\documentclass[letter]{ptptex}

\usepackage{graphicx}
\usepackage{wrapft}

\def\la{\mathrel{\mathpalette\fun <}}

\def\fun#1#2{\lower3.6pt\vbox{\baselineskip0pt\lineskip.9pt
\ialign{$\mathsurround=0pt#1\hfil##\hfil$\crcr#2\crcr\sim\crcr}}}

\newcommand{\beq}{\begin{equation}}
\newcommand{\eeq}{\end{equation}}
\newcommand{\bea}{\begin{eqnarray}}
\newcommand{\eea}{\end{eqnarray}}

\newcommand{\rra}{\right\rangle}
\newcommand{\lla}{\left\langle}
\newcommand{\bfi}[1]{\mbox{\boldmath $#1$}}

\newcommand{\vb}{{\bfi b}}
\newcommand{\vk}{{\bfi k}}

\newcommand{\vrr}{{\bfi r}}
\newcommand{\vR}{{\bfi R}}

\title{
Eikonal Reaction Theory for Neutron-Removal Reaction
}

\author{Masanobu \textsc{YAHIRO}, Kazuyuki \textsc{OGATA}, and Kosho \textsc{MINOMO}}
\inst{Department of Physics, Kyushu University, Fukuoka 812-8581, Japan}

\recdate{\today}

\abst{
We present an accurate method of treating
the one-neutron removal reaction at intermediate incident energies
induced by both nuclear and Coulomb interactions.
In the method, the nuclear and Coulomb breakup processes
are consistently treated by the method of
continuum discretized coupled channels
without making the adiabatic approximation to the Coulomb interaction,
so that the removal cross section calculated never diverges.
This method is applied to recently measured one-neutron removal cross section
for $^{31}$Ne+$^{12}$C scattering at 230~MeV/nucleon and
$^{31}$Ne+$^{208}$Pb scattering at 234~MeV/nucleon.
The spectroscopic factor and the asymptotic normalization coefficient
of the last neutron in $^{31}$Ne are evaluated.
}

\begin{document}

\maketitle

\section{Introduction}

Unstable nuclei have exotic properties such as
the halo structure~\cite{Tanihata,Jensen,Jonson}
and the change of magicity for nuclei in the region 
called 
``Island of inversion"~\cite{Warburton,Caurier,Utsuno}. 
These novel quantum properties have inspired
extensive experimental and theoretical studies.
An important experimental tool of exploring such exotic properties is
the nucleon removal reaction; see for example Ref.~\citen{Gade}.
The theoretical tool of analyzing such an inclusive reaction
is the Glauber model~\cite{Glauber}.
The theoretical foundation of the model is investigated
in Ref.~\citen{Yahiro-Glauber}.
The model is based on the eikonal and the adiabatic approximation. 
It is well known that the adiabatic approximation makes
the removal cross section diverge when 
the Coulomb interaction is included. 
The Glauber model has thus been applied mainly for
lighter targets in which the Coulomb interaction is
negligible; see for example 
Refs.~\citen{Hussein,Hencken,Gade,Ogawa01,Tostevin,Bertulani-92,Bertulani-04}. 
Very recently, the Glauber model with the Coulomb correction 
was proposed~\cite{Ibrahim,Capel-08}. 
In the Coulomb-corrected eikonal model (CCE)~\cite{Ibrahim,Capel-08}, 
the E1 (dipole) component of the eikonal Coulomb phase is replaced 
by that based on the first-order perturbation. 
For the elastic-breakup scattering of $^{11}$Be from a $^{208}$Pb target at 
69~MeV/nucleon, it was shown in Ref.~\citen{Capel-08} that 
CCE well simulates  results of the dynamical eikonal 
approximation (DEA)~\cite{DEA}.

Meanwhile, the method of continuum discretized coupled channels
(CDCC)~\cite{CDCC-review1,CDCC-review2}
is an accurate method of treating
exclusive reactions such as the elastic scattering and
the elastic breakup reaction in which
the target is not excited whereas
the projectile is broken up into its fragments. 
The theoretical foundation of CDCC was shown in 
Refs.~\citen{CDCC-foundation1,CDCC-foundation2,CDCC-foundation3}. 
Actually, CDCC has succeeded in reproducing data on the scattering of 
not only stable nuclei but also unstable nuclei; 
see for example Refs.~\citen{Tostevin2,Davids,Matsumoto3,Matsumoto4,THO-CDCC,4body-CDCC-bin,Matsumoto:2010mi} and references therein. 
The dynamical eikonal approximation (DEA)~\cite{DEA}
is also an accurate method of 
treating exclusive reactions at intermediate and high incident energies 
where the eikonal approximation is reliable. 
The nucleon removal reaction is composed of the exclusive
elastic-breakup component and
the inclusive nucleon-stripping component. 
CDCC and DEA can evaluate the elastic-breakup cross section, but not
the stripping cross section.

The experimental exploration of halo nuclei
is moving from lighter nuclei such
as He and C isotopes to relatively heavier nuclei such as Ne isotopes.
Very recently, a halo structure of $^{31}$Ne has been reported by the
experiment on the one-neutron removal reaction $\sigma_{-n}$ 
at 230~MeV/nucleon
not only for a $^{12}$C target but also for a $^{208}$Pb
target~\cite{Nakamura}.
This is the heaviest halo nucleus in the present
stage confirmed experimentally and also reside
within the region of ``Island of inversion".
The determination of the spin-parity of $^{31}$Ne
in its ground state is essential to understand the nature
of ``Island of inversion". 
The one-neutron removal reaction has been analyzed with 
the Glauber model~\cite{30Ne}; for a $^{208}$Pb target, 
the elastic breakup component due to one-step E1 transition
is added.

In this Letter, we present an accurate method of treating
the one-neutron removal reaction at intermediate incident energies such
as 200~MeV/nucleon induced by both nuclear and Coulomb interactions.
In the method, the nuclear and Coulomb breakup processes
are accurately treated by CDCC without making the adiabatic approximation
to the latter,
so that the removal cross section calculated never diverges
and reliable even in the presence of the Coulomb interaction.
Thus, this method is an essential extension of
the Glauber theory and CDCC. 
This method is applied to
the one-neutron removal reaction
of the $^{31}$Ne+$^{12}$C scattering at 230~MeV/nucleon and
the $^{31}$Ne+$^{208}$Pb scattering at 234~MeV/nucleon. 
The results of the present method are compared with those of CCE. 
The spectroscopic factor ${\cal S}$ and
the asymptotic normalization coefficient $C_{\rm ANC}$~\cite{Xu}
of the last neutron in $^{31}$Ne are evaluated with the new theory.

\section{Formulation}
We consider the one-neutron halo nucleus as a projectile (P) and
take a single-particle model for the nucleus;
namely, the projectile consists of a core nucleus (c) and a neutron (n).
The scattering of P on a target (T) is then described by
the three-body (c+n+T) Schr\"odinger equation
\beq
   \left[ -\frac{\hbar^2}{2\mu}\nabla_R^2+ h + U(r_{\rm c},r_{\rm n})-E \right]\Psi=0
\label{Schrodinger-eq}
\eeq
with the interaction
\beq
 U = U_{\rm n}^{\rm (N)}(r_{\rm n})
 + U_{\rm c}^{\rm (N)}(r_{\rm c}) + U_{\rm c}^{\rm (C)}(r_{\rm c}) ,
\label{Pot}
\eeq
where $\mu$ is the reduced mass between P and T.
The three-dimensional vector $\vR=(\vb,Z)$ stands
for the coordinate between P and T, whereas
$\vrr_{\rm x}$ (x=n or c) is the coordinate between x and A
and $\vrr$ means the coordinate between c  and n.
The operator $h=T_r +V(\vrr)$ is the projectile Hamiltonian composed of
the kinetic-energy operator $T_r$ and the interaction $V(\vrr)$.
The potential $U_{\rm x}^{\rm (N)}$ is
the nuclear part of the optical potential between x and T,
whereas $U_{\rm c}^{\rm (C)}$ is the Coulomb interaction
between c and T.

Now, we consider scattering at intermediate incident energies,
say 200~MeV/nucleon, for which the eikonal approximation is quite good.
The three-body wave function $\Psi$ is first assumed to be
\bea
    \Psi={\hat O} \psi(\vR,\vrr) ,
\label{product}
\eea
where the operator ${\hat O}$ is defined by
\bea
{\hat O} =\frac{1}{\sqrt{\hbar {\hat v}}} e^{i {\hat K} \cdot Z}
\eea
with the wave-number operator ${\hat K}=\sqrt{2\mu(E-h)}/{\hbar}$
and the velocity operator ${\hat v}={\hbar {\hat K}}/{\mu}$
of the relative motion between P and T.
When \eqref{product} is inserted into \eqref{Schrodinger-eq},
we have three terms ${\hat O}[\nabla_R^2\psi]$,
$[\nabla_R{\hat O}] \cdot [\nabla_R\psi]$
and $[\nabla_R^2 {\hat O}]\psi$.
The first term is much smaller than the others,
since $\psi$ is slowly varying with $\vR$ compared
with ${\hat O}$. Neglecting the first term leads to
\bea
  i \frac{d\psi}{dZ}={\hat O^{\dagger}}U{\hat O}\psi .
\label{eikonal-eq}
\eea
Regarding $Z$ as ``time" virtually and solving \eqref{eikonal-eq}
iteratively, we obtain the formal solution
\bea
\psi=\exp\Big[
-i{\cal P}\int_{-\infty}^Z dZ' {\hat O^{\dagger}}U{\hat O}
\Big] ,
\label{WF}
\eea
where ${\cal P}$ is the ``time" ordering operator. Taking $Z$ to $\infty$
in \eqref{WF}, we finally get the $S$-matrix operator
\bea
S=\exp\Big[
-i{\cal P}\int_{-\infty}^{\infty} dZ {\hat O^{\dagger}}U{\hat O}
\Big]  .
\label{S-matrix-operator}
\eea
In the Glauber model, the adiabatic approximation is made as the secondary
approximation. In the approximation, $h$ is replaced by the ground-state
energy $\epsilon_0$, and hence ${\hat O^{\dagger}}U{\hat O}$ and
${\cal P}$ in \eqref{S-matrix-operator} are reduced to
$U/(\hbar v_0)$ and 1, respectively, where $v_0$ is the velocity of P
in the ground state relative to T.
This is nothing but the $S$-matrix in the Glauber model.

At intermediate incident energies,
the adiabatic approximation is good for the short-range nuclear interactions,
$U_{\rm n}^{\rm (N)}$ and $U_{\rm c}^{\rm (N)}$, but not for the
long-range Coulomb interaction
$U_{\rm c}^{\rm (C)}$. This can be understood from the matrix element
\beq
\int_{-\infty}^{\infty} dZ
\langle \varphi_{\vk} | {\hat O^{\dagger}}U{\hat O} | \varphi_{0} \rangle
\approx
\frac{ e^{i (K_0 - K) R_U} }{\hbar v_0}
\int_{-\infty}^{\infty} dZ
\langle \varphi_{\vk} | U | \varphi_{0} \rangle
\label{Estimation}
\eeq
between the ground state $\varphi_{0}$ of P with the intrinsic
energy $\epsilon_0$
and the continuum state $\varphi_{\vk}$ of P with the intrinsic momentum and
energy, $\hbar \vk$ and $\epsilon(k)$.
In (\ref{Estimation}), $\hbar K_0$ ($\hbar K$)
is the momentum of P in the ground (continuum) state relative to T,
and $R_U$ is the range of the interaction considered.
For the $^{31}$Ne
breakup reaction at 200~MeV/nucleon, the typical excitation energy
$\epsilon(k) \approx 1$~MeV.
For the Coulomb interaction, because of its long-range property,
$(K_0 - K) R_U$ is large even if $K_0 - K$ is small.
This means that we can not set $K_0 - K$ to zero, that is,
the adiabatic approximation does not work.
Actually, the elastic-breakup cross section diverges
in the adiabatic limit of $K=K_0$ because of
the slow decrease of the matrix element \eqref{Estimation}
in $b$~\cite{Eikonal-CDCC,Capel-08}.
For the nuclear interactions, meanwhile, $R_U \approx 11$~fm and hence
$(K_0 - K) R_U \approx 0.074$. Thus, the adiabatic
approximation is good for the nuclear interactions.
This indicates that $U_{\rm n}^{\rm (N)}$ is commutable with
${\hat O}$ with high accuracy. Therefore, we can take the replacement
\beq
    {\hat O^{\dagger}}U_{\rm n}^{\rm (N)} {\hat O} \rightarrow
     U_{\rm n}^{\rm (N)}/(\hbar v_0) .
\label{replacement-N}
\eeq
The replacement \eqref{replacement-N} is accurate as shown later by
numerical calculations. Using this replacement, one can get 
the important result
\beq
  S = S_{\rm n}S_{\rm c}
  \label{S-separation}
\eeq
with 
\bea
  S_{\rm n}&=&
    \exp\Big[   - \frac {i}{\hbar v_0} \int_{-\infty}^{\infty} dZ 
    U_{\rm n}^{\rm (N)} \Big] ,
  \label{Sn} \\
  S_{\rm c}&=&\exp\Big[
-i{\cal P}\int_{-\infty}^{\infty} dZ {\hat O^{\dagger}}(U_{\rm c}^{\rm (N)}+U_{\rm c}^{\rm (C)})
{\hat O} \Big] .
\label{Sc}
\eea
Equation \eqref{S-separation} can be derived as follow. 
The replacement \eqref{replacement-N} leads \eqref{eikonal-eq} to 
\bea
  i \frac{d\psi}{dZ}={\hat O^{\dagger}}U_{\rm c}{\hat O}\psi 
  +\frac{1}{\hbar v_0} U_{\rm n}^{\rm (N)}\psi
\label{eikonal-eq-2}
\eea
with $U_{\rm c}=U_{\rm c}^{\rm (N)} + U_{\rm c}^{\rm (C)}$. 
Defining $\xi$ as $\psi=Q\xi$ with 
\bea
Q=\exp\Big[   - \frac {i}{\hbar v_0} \int_{-\infty}^{Z} dZ' 
    U_{\rm n}^{\rm (N)} \Big] 
\eea
and inserting $\psi=Q\xi$ into \eqref{eikonal-eq-2} leads to 
\bea
  i \frac{d\xi}{dZ}={\hat O^{\dagger}}U_{\rm c}{\hat O}\xi, 
\label{eikonal-eq-3}
\eea
where use has been made of the fact that $Q$, i.e. $U_{\rm n}^{\rm (N)}$, 
is commutable with ${\hat O}$ with high accuracy. 
The formal solution of \eqref{eikonal-eq-3} is $S_{\rm c}$, 
whereas $S_{\rm n}$ is obtained from $Q$ by taking $Z\to\infty$. 
Noting that $\psi=Q\xi$, one can reach \eqref{S-separation}.

Thus, $S$ can be separated into the neutron part $S_{\rm n}$ and
the core part $S_{\rm c}$.
We can not calculate $S_{\rm c}$ directly with \eqref{Sc},
because it includes the operators ${\hat O}$ and ${\cal P}$.
However, $S_{\rm c}$ is the approximate solution of the Schr\"odinger equation
\beq
   \left[ -\frac{\hbar^2}{2\mu}\nabla_R^2 + h + U_{\rm c}^{\rm (N)}(r_{\rm c})
   + U_{\rm c}^{\rm (C)}(r_{\rm c})-E \right]\Psi_c=0, 
\label{Schrodinger-eq-core}
\eeq
when the eikonal approximation is made. We can calculate the matrix elements
$\langle \varphi_{0} | S_{\rm c} | \varphi_{0} \rangle$ and
$\langle \varphi_{\vk} | S_{\rm c} | \varphi_{0} \rangle$ by solving
\eqref{Schrodinger-eq-core} with eikonal-CDCC~\cite{Eikonal-CDCC} in 
which the eikonal approximation is made in the framework of CDCC. 
Non-eikonal corrections to $S_{\rm c}$ can be easily made by solving 
\eqref{Schrodinger-eq-core} with CDCC instead of eikonal-CDCC, 
although it is not necessary for the present intermediate scattering. 
As mentioned above, $S_{\rm n}$ is obtained by \eqref{Sn}.

The reaction theory constructed above is referred to as
the eikonal reaction theory (ERT) in this Letter.
We can test the accuracy of the eikonal approximation
by comparing the $S$-matrix elements of CDCC and
eikonal-CDCC~\cite{Eikonal-CDCC}.
We confirmed that the approximation is quite good for the present
scattering system. This approximation is good even for
lower incident energies such as 30~MeV/nucleon~\cite{Eikonal-CDCC}.
So all CDCC calculations in this Letter
are done with eikonal-CDCC.

We can derive several kinds of cross sections with the product form
\eqref{S-separation}, following the formulation on the cross sections in
the Glauber model~\cite{Hussein,Hencken}.
The one-neutron removal cross section
$\sigma_{-n}$ is the sum of the total elastic breakup (diffractive)
cross section $\sigma_{\rm bu}$ and
the total neutron-stripping cross section $\sigma_{\rm str}$
in which n is absorbed by T:
\beq
\sigma_{-n}=\sigma_{\rm bu}+\sigma_{\rm str}
\eeq
with
\bea
\sigma_{\rm str}
&=&\int d^2 \vb
\langle \varphi_{0} | |S_{\rm c}|^2(1-|S_{\rm n}|^2) | \varphi_{0} \rangle
\nonumber \\
&=& [\sigma_{\rm R}-\sigma_{\rm bu}]
-[\sigma_{\rm R}(-n)-\sigma_{\rm bu}(-n)],
\label{one-removal-Xsec}
\eea
where $\sigma_{\rm R}$ and $\sigma_{\rm bu}$ are
the total reaction and elastic-breakup cross sections, respectively,
defined by
\bea
\sigma_{\rm R}&=&
\int d^2 \vb [1-
| \langle \varphi_{0} | S_{\rm c}S_{\rm n} | \varphi_{0} \rangle|^2] , \\
\sigma_{\rm bu}&=&
\int d^2 \vb [
\langle \varphi_{0} | |S_{\rm c}S_{\rm n}|^2 | \varphi_{0} \rangle -
|\langle \varphi_{0} | S_{\rm c}S_{\rm n} | \varphi_{0} \rangle|^2] ,~~~
\eea
and
$\sigma_{\rm R}(-n)$ and $\sigma_{\rm bu}(-n)$ correspond to
the total reaction and elastic-breakup cross sections, respectively,
in which
$S_{\rm c}S_{\rm n}$ is replaced by $S_{\rm c}$.
The last form of \eqref{one-removal-Xsec} means that
$\sigma_{-n}$ can be obtained from
$\sigma_{\rm R}$, $\sigma_{\rm bu}$, $\sigma_{\rm R}(-n)$
and $\sigma_{\rm bu}(-n)$ by solving
\eqref{Schrodinger-eq} and \eqref{Schrodinger-eq-core}
with CDCC.

\section{Model setting}
We take the same three-body model as in Ref.~\citen{30Ne}
except for minor difference,
since the Glauber model calculation based on the three-body model
yields a reasonable value of $\sigma_{\rm -n}$.
The nuclear potential $U_{\rm x}^{\rm (N)}$ (x=n or c) is
calculated by the folding model in which the effective
nucleon-nucleon (NN) interaction $t$ is folded with the densities
of particles T and x.
As for $t$, we take the parameter set in
Refs.~\citen{30Ne,NNint} for $E_{\rm NN}=240$~MeV,
where $E_{\rm NN}$ is the energy
of NN scattering
\footnote{
In Ref.~\citen{NNint}, the effective interaction is presented as the
profile function $\Gamma(b_{\rm NN})$. It is a function of not
the distance $r_{\rm NN}$ between two nucleons but
the impact parameter $b_{\rm NN}$.
The interaction $t(r_{\rm NN})$ is so constructed that
the profile function obtained from $t(r_{\rm NN})$ can agree with
that of Ref.~\citen{NNint}.
}.
The folding potential reproduces the reaction cross section of
$^{12}$C + $^{12}$C scattering in the wide range of incident
energies~\cite{NNint}. In Ref.~\citen{30Ne},
the $^{30}$Ne density is calculated by the single-particle model
in which the separation energies of last neutron and proton reproduce
the experimental values $3$~MeV and $2.4$~MeV, respectively;
for the potential parameters, see the case of $(r_0,a)=(1.25,0.75)$
in Table I of Ref.~\citen{30Ne},
where $r_0$ and $a$ are
the radius and the diffuseness parameters of the Woods-Saxon form in units
of fm. The target density is evaluated by the Hartree-Fock calculation with
the effective NN interaction Gogny-D1S~\cite{Gogny,D1S}.
The $^{30}$Ne-n potential $V$ is determined so that
the neutron separation energy $B_{\rm n}$ becomes 0.33~MeV;
we take the case of
$(r_0,a)=(1.25,0.75)$ in Table II of Ref.~\citen{30Ne}
for $1p3/2$ and the case of $(r_0,a)=(1.25,0.70)$ for $0f7/2$.
A large $\sigma^{\rm exp}_{\rm -n}$ indicates
that $^{31}{\rm Ne}$ is a one-neutron halo nucleus. This structure
is realized with small angular momenta between c and n. However,
the possibility of the $2s1/2$ orbit is small from the theoretical point of
view, since the single-particle energy of $2s1/2$ is much higher than those of
$1p3/2$ and $0f7/2$. Hence, we consider the $1p3/2$ and $0f7/2$ orbits.

\section{Results}
In ERT, the adiabatic approximation is assumed to be good for
the nuclear potential $U_{\rm n}^{\rm (N)}$.
This can be tested by setting
$U=U_{\rm n}^{\rm (N)}(r_n)+U_{\rm c}^{\rm (C)}(R)+U_{\rm c}^{\rm (N)}(R)$
in the Schr\"odinger equation \eqref{Schrodinger-eq}.
In this setup,
the projectile breakup is induced only by $U_{\rm n}^{\rm (N)}(r_n)$, since
the argument $r_{\rm c}$ of $U_{\rm c}^{\rm (C)}$ and $U_{\rm c}^{\rm (N)}$
is replaced by $R$.
Switching the adiabatic approximation on in the Schr\"odinger equation
corresponds to the replacement \eqref{replacement-N}.
For the scattering of $^{31}$Ne($1p3/2$) from $^{208}$Pb,
the error due to the approximation
is 0.2\% for $\sigma_{\rm R}$, 1.9\% for
$\sigma_{\rm bu}$, 4.1\% for $\sigma_{\rm str}$ and
3.3\% for $\sigma_{\rm -n}$. Thus, the error is small.

Table \ref{table1} presents several kinds of cross sections and
the spectroscopic factor
${\cal S}=\sigma^{\rm exp}_{\rm -n}/\sigma^{\rm th}_{\rm -n}$
for $^{12}$C and $^{208}$Pb targets.
As an important result,
${\cal S}[1p3/2]=$0.693 (0.682) for $^{12}$C ($^{208}$Pb),
whereas ${\cal S}[0f7/2]=$2.47 (5.65) for $^{12}$C ($^{208}$Pb).
Thus, ${\cal S}[1p3/2]$ little depends on the target and less than 1,
but ${\cal S}[0f7/2]$ does not satisfy these conditions.
In Ref.~\citen{Nakamura}, the Coulomb component
of $\sigma_{\rm -n}[^{208}{\rm Pb}]$ for a $^{208}$Pb target
is estimated to be 540~mb from the experimental values of
$\sigma_{\rm -n}[^{208}{\rm Pb}]$ and $\sigma_{\rm -n}[^{12}{\rm C}]$.
In ERT, the Coulomb component
of $\sigma_{\rm -n}[^{208}{\rm Pb}]$ agrees with
$\sigma_{\rm bu}[^{208}{\rm Pb}]$ with good accuracy.
The spectroscopic factor evaluated from the Coulomb component is
${\cal S'}=540/\sigma^{\rm th}_{\rm bu}=0.675$ for the $1p3/2$ orbit and
7.36 for the $0f7/2$ orbit. Thus, ${\cal S'}$ is consistent
with ${\cal S}$ only for the $1p3/2$ orbit.
Hence, we can infer that
the major component of the $^{31}{\rm Ne}_{\rm g.s.}$ wave function
is $^{30}{\rm Ne}(0^+) \otimes 1p3/2$ (${\cal S} \sim 0.69$).
We adopt in the following this configuration.

\begin{table}
\begin{center}
\caption
{Several kinds of cross sections and the spectroscopic factors for
$^{31}$Ne+$^{12}$C scattering at 230~MeV/nucleon and
$^{31}$Ne+$^{208}$Pb scattering at 234~MeV/nucleon.
The cross sections are presented in units of mb and the data are taken from
Ref.~\citen{Nakamura}.
}
\label{table1}
\begin{tabular}{cccccccc}
\hline \hline
 & \multicolumn{3}{c}{$^{12}$C target} & \hspace{2mm} &
 \multicolumn{3}{c}{$^{208}$Pb target} \\ \cline{2-4}  \cline{6-8}
 {} & ~~~$p_{3/2}$~~~ & ~~~$f_{7/2}$~~~ & ~~~exp~~~ &
 {} & ~~~$p_{3/2}$~~~ & ~~~$f_{7/2}$~~~ & ~~~exp~~~ \\ \hline
 $\sigma_{\rm R}$ & 1572.5 & 1489.9 & & & 5518.0 & 4589.5 & \\
 $\sigma_{\rm bu}$ & 23.3 & 3.3 & & & 799.5 & 73.0 & (540) \\
 $\sigma_{\rm R}$(-n) & 1463.5 & 1458.6 & & & 5151.5 & 4524.2 & \\
 $\sigma_{\rm bu}$(-n) & 4.5 & 1.0 & & & 677.2 & 60.5 & \\ \hline
 $\sigma_{\rm str}$ & 90 & 29 & & & 244 & 53 & \\
 $\sigma_{\rm -n}$ & 114 & 32 & 79 & & 1044 & 126 & 712 \\
 ${\cal S}$ & 0.693 & 2.47 & & & 0.682 & 5.65 & \\ \hline \hline
\end{tabular}
\end{center}
\end{table}

In Ref.~\citen{30Ne}, ${\cal S}$ is calculated with the Glauber model; 
for a $^{208}$Pb target, the elastic breakup component 
due to one-step E1 transition is added. 
The resulting spectroscopic factor for the $1p3/2$ orbit is
${\cal S}=0.822$ for a $^{12}$C target and 0.624
for a $^{208}$Pb target~\cite{30Ne}. 
The difference between the results of the present study and Ref.~\citen{30Ne}
for $^{12}$C mainly comes from the neglect of $\sigma_{\rm bu}$ in the latter, 
and that for $^{208}$Pb seems to stem from Coulomb higher-order contributions 
to $\sigma_{\rm bu}$. 
Neglect of the Coulomb interaction in $\sigma_{\rm str}$
may also be responsible for the difference.
Thus, CCE is fairly good also for the present system.

The potential $V$ between c and n is not well known. Hence, ${\cal S}$ has
a theoretical error coming from the potential ambiguity. The error
is often estimated by changing each of $r_0$ and $a$ by 30\%.
When the one-neutron separation energy $B_{\rm n}$ of $^{31}$Ne is 0.33~MeV,
${\cal S}= 0.693 \pm 0.133 \pm 0.061$
for a $^{12}$C target and
$0.682 \pm 0.133 \pm 0.062$
for a $^{208}$Pb target, where the second and third numbers
following the mean value
stand for the theoretical and experimental uncertainties, respectively.
Thus, ${\cal S}$ includes a sizable theoretical error.
This situation completely changes if we look at the
asymptotic normalization coefficient (ANC).
For $r$ out of the range of $V$, ANC $C_{\rm ANC}$ is defined by~\cite{Xu}
\beq
I_{lj}(r) = C_{\rm ANC} h_l^{(+)}(i \kappa r)
\eeq
with the radial part of the overlap function
$I_{lj}(r)=\lla \phi_{\rm c} \phi_{\rm n}
| \phi_{\rm P} \rra$, where
$\phi_{\rm c}$ and $\phi_{\rm n}$ are
the intrinsic states of c and n, respectively, whereas $\phi_{\rm P}$ is
the ground state of P. $h_{l}^{(+)}$ is a spherical Hankel function and
$\kappa$ is the relative wave number
between c and n in the ground state of P.
For large $r$, $I_{lj}(r)$ is related to the normalized single-particle
wave function
$\varphi_{lj}(r)$ that is determined from $V$ as
\bea
I_{lj}(r) = \sqrt{\cal S} \varphi_{lj}(r) = \sqrt{\cal S}
C_{\rm ANC}^{({\rm sp})} h_l^{(+)}(i \kappa r)
\eea
with the single-particle ANC $C_{\rm ANC}^{({\rm sp})}$. Hence, we have
\bea
C_{\rm ANC} = \sqrt{\cal S} C_{\rm ANC}^{({\rm sp})}.
\eea
When $B_{\rm n}=0.33$~MeV,
$C_{\rm ANC}= 0.320 \pm 0.010 \pm 0.028$~fm$^{-1/2}$
for a $^{12}$C target and
$0.318 \pm 0.008 \pm 0.029$~fm$^{-1/2}$
for a $^{208}$Pb target.
Thus, $C_{\rm ANC}$  has a much smaller theoretical error than
${\cal S}$. This means that the one-nucleon removal reaction is quite
peripheral.

\begin{figure}[htbp]
\begin{center}
 \includegraphics[width=0.5\textwidth,clip]{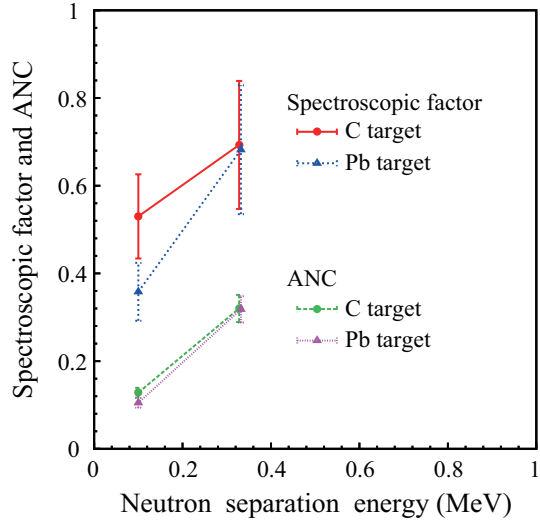}
 \caption{(Color online) $B_{\rm n}$ dependence of ${\cal S}$ and $C_{\rm ANC}$.
 The error bar contains theoretical and experimental errors.
 }
 \label{Fig-Sn}
\end{center}
\end{figure}

The experimental value of $B_{\rm n}$ is not precisely
determined: $B_{\rm n}=0.29 \pm 1.64$~MeV~\cite{Sn}.
It is thus better to see $B_{\rm n}$ dependence of
$C_{\rm ANC}$ and ${\cal S}$.
When $B_{\rm n}=0.1$~MeV,
$C_{\rm ANC}= 0.128 \pm 0.003 \pm 0.011$~fm$^{-1/2}$
and
${\cal S}= 0.530 \pm 0.084 \pm 0.047 $
for a $^{12}$C target, and
$C_{\rm ANC}= 0.105 \pm 0.004 \pm 0.010$~fm$^{-1/2}$
and
${\cal S}= 0.358 \pm 0.057 \pm 0.033 $
for a $^{208}$Pb target.
These values are plotted in Fig.~\ref{Fig-Sn}.
$C_{\rm ANC}$ and ${\cal S}$ are sensitive to the value of $B_{\rm n}$.
We can see from $B_{\rm n}$ dependence of ${\cal S}$ for a $^{208}$Pb target
that ${\cal S} < 1$ when $B_{\rm n} \la 0.6$~MeV.
It is thus necessary to determine $B_{\rm n}$ experimentally in future
in order to evaluate $C_{\rm ANC}$ and ${\cal S}$ properly.
However, we can say at least that $C_{\rm ANC}$ has
a smaller theoretical error and weaker target dependence than
${\cal S}$ for any value of $B_{\rm n}$.

\section{Summary}
We present an accurate method of treating
the one-neutron removal reaction at intermediate energies induced by
both the short-range nuclear and the long-range Coulomb interaction.
In the theory called ERT,
the nuclear and Coulomb breakup processes are accurately and
consistently treated by CDCC
without making the adiabatic approximation to the latter,
so that the removal cross section never diverges and hence reliable
even in the presence of the Coulomb interaction.
This method is applicable also for other inclusive reactions such as
the two-nucleon removal reaction induced
by nuclear and Coulomb interactions. 

ERT is a theory of presenting the Schr\"odinger equation
\eqref{Schrodinger-eq-core} to calculate inclusive cross sections. 
When these equations are solved, any accurate method is useful. 
For intermediate and high incident energies, 
one can use DEA instead of eikonal-CDCC. 
For lower incident energies where 
the eikonal approximation is not perfectly accurate, 
one should make non-eikonal corrections to inclusive cross sections. 
This can be done easily by using CDCC instead of eikonal-CDCC.

$C_{\rm ANC}$ and ${\cal S}$ of the last neutron in $^{31}$Ne
are evaluated from the one-neutron removal reaction
of the $^{31}$Ne+$^{12}$C scattering at 230~MeV/nucleon and
the $^{31}$Ne+$^{208}$Pb scattering at 234~MeV/nucleon.
$C_{\rm ANC}$ has a smaller theoretical error and
weaker target-dependence than ${\cal S}$. Thus, $C_{\rm ANC}$ is determined
more accurately than ${\cal S}$.
This may change the future strategy for the spectroscopy of
unstable nuclei.

When the last neutron in $^{31}$Ne is in the $1p3/2$ orbit,
${\cal S} < 1$ for $B_{\rm n} \la 0.6$~MeV, and ${\cal S}$ and $C_{\rm ANC}$
have weaker target dependence.
When the last neutron is in the $1f7/2$ orbit, meanwhile,
${\cal S} > 1$ and ${\cal S}$ and $C_{\rm ANC}$
have stronger target dependence.
These results indicate that the last neutron is mainly in the $1p3/2$ orbit
and imply that $^{31}$Ne is deformed.
This sort of deformation is expected
as an origin of the change of magicity~\cite{Warburton,Caurier,Utsuno}.

\section*{Acknowledgements}

M.Y. thanks D. Baye, P. Descouvemont and Y. Suzuki for useful discussions. 
The numerical calculations of this work were performed
on the computing system in Research Institute
for Information Technology of Kyushu University.

%



\begin{thebibliography}{99}

  
\bibitem{Tanihata}
  I.~Tanihata, \textit{et al}.,
  \newblock
  Phys.\ Lett.\ B {\bf 289} (1992), 261. \\
  I.~Tanihata, \newblock
   J.\ Phys.\ G {\bf 22} (1996), 157.

\bibitem{Jensen}
  A.~S.~Jensen, \textit{et al}., \newblock
    Rev.\ Mod.\ Phys.\ {\bf 76} (2004), 215.

\bibitem{Jonson}
   B. Jonson, \newblock
   Phys.\ Rep.\ {\bf 389} (2004), 1.

\bibitem{Warburton}
  E.~K.~Warburton, J.~A.~Becker, and B.~A.~Brown, Phys.\
    Phys.\ Rev.\ C {\bf 41} (1990), 1147.

\bibitem{Caurier}
  E.~Caurier, \textit{et al.},
  Phys.\ Rev.\ C {\bf 58} (1998), 2033.

\bibitem{Utsuno}
  Y.~Utsuno, \textit{et al.},
  Phys.\ Rev.\ C {\bf 60} (1999), 054315.

\bibitem{Gade}
  A.~Gade, \textit{et al.},
  Phys.\ Rev.\ C {\bf 77} (2008), 044306.

\bibitem{Glauber}
    R.J.~Glauber, {\it in Lectures in Theoretical
    Physics} (Interscience, New York, 1959), Vol. 1, p.315.

\bibitem{Yahiro-Glauber}
M.~Yahiro, K.~Minomo, K.~Ogata, and M.~Kawai,
  Prog.\ Theor.\ Phys.\ {\bf 120} (2008), 767.


\bibitem{Hussein}
M. S. Hussein and K. W. McVoy,
Nucl. Phys. A \textbf{445} (1985), 124.

\bibitem{Bertulani-92}
C. A. Bertulani and K. W. McVoy, Phys. Rev. C {\bf 46} (1992), 2638. 

\bibitem{Hencken}
K. Hencken, G. Bertsch, and H. Esbensen,
Phys. Rev. C \textbf{54} (1996), 3043.


\bibitem{Tostevin}
J.S. Al-Khalili and J.A. Tostevin, Phys. Rev. Lett. {\bf 76} (1996), 3903;
J.A. Tostevin and B.A. Brown, Phys. Rev. C {\bf 74} (2006), 064604. 

\bibitem{Ogawa01}
K. Yabana, Y. Ogawa, and Y. Suzuki, Nucl. Phys. A {\bf 539} (1992), 295;
Y. Ogawa, K. Yabana, and Y. Suzuki, Nucl. Phys. A {\bf 543} (1992), 722;
Y. Ogawa, T. Kido, K. Yabana, and Y. Suzuki,
Prog. Theor. Phys. Supplement {\bf 142} (2001), 157. 


\bibitem{Bertulani-04}
C. A. Bertulani and P. G. Hansen, Phys. Rev. C {\bf 70} (2004), 034609. 



\bibitem{Ibrahim}
B. Abu-Ibrahim and Y. Suzuki,
Prog. Theor. Phys. {\bf 112} (2004), 1013;
B. Abu-Ibrahim and Y. Suzuki,
Prog. Theor. Phys. {\bf 114} (2005), 901.



\bibitem{Capel-08}
    P. Capel, D. Baye, and Y. Suzuki,
    Phys.\ Rev.\ C {\bf 78} (2008), 054602.


\bibitem{DEA}
D. Baye, P. Capel, and G. Goldstein, 
Phys. Rev. Lett. {\bf 95} (2005), 082502;
G. Goldstein, D. Baye, and P. Capel, 
Phys. Rev. C {\bf 73} (2006), 024602.



\bibitem{CDCC-review1}
  M.~Kamimura, 
M.~Yahiro, Y.~Iseri, Y.~Sakuragi, H.~Kameyama and
M.~Kawai, \newblock
  Prog.\ Theor.\ Phys.\ Suppl.\ {\bf 89} (1986), 1.

\bibitem{CDCC-review2}
  N.~Austern, 
Y.~Iseri, M.~Kamimura, M.~Kawai, G.~Rawitscher and
M.~Yahiro, \newblock
  Phys.\ Rep.\ {\bf 154} (1987), 125. 


\bibitem{CDCC-foundation1}
N.~Austern, M.~Yahiro, and M.~Kawai,
\newblock
Phys.\ Rev.\ Lett. {\bf 63} (1989), 2649.

\bibitem{CDCC-foundation2}
N.~Austern, 
M.~Kawai, and  M.~Yahiro, \newblock
Phys.\ Rev.\ C {\bf 53} (1996), 314.

\bibitem{CDCC-foundation3}
A. Deltuva, A.M.~Moro, E.~Cravo, F.M.~Nunes, and A.C.~Fonseca,
Phys.\ Rev.\ C {\bf 76} (2007), 064602.
  

\bibitem{Tostevin2}
J.A. Tostevin, F.M. Nunes, and I.J. Thompson,
Phys. Rev. C {\bf 63} (2001), 024617.

\bibitem{Davids}
B. Davids, S.M. Austin, D. Bazin, H. Esbensen,
B.M. Sherrill, I.J. Thompson, and J.A. Tostevin,
Phys. Rev. C {\bf 63} (2001), 065806.


\bibitem{Matsumoto3}
T.~Matsumoto, 
E.~Hiyama, K.~Ogata, Y.~Iseri, M.~Kamimura,
S.~Chiba, and M.~Yahiro,
Phys.\ Rev.\ C {\bf 70} (2004), 061601(R).


\bibitem{Matsumoto4}
T.~Matsumoto, 
T.~Egami, K.~Ogata, Y.~Iseri, M.~Kamimura, and M.~Yahiro,
Phys.\ Rev.\ C {\bf 73} (2006), 051602(R).



\bibitem{THO-CDCC}
M.~Rodr\'{i}guez-Gallardo, 
J.~M.~Arias, J.~G\'{o}mez-Camacho,
R.~C.~Johnson, A.~M.~Moro, I.~J.~Thompson, and J.~A.~Tostevin,
\newblock
Phys.\ Rev.\ C {\bf 77} (2008), 064609.

\bibitem{4body-CDCC-bin}
M.~Rodr\'{i}guez-Gallardo,
J. M. Arias, J. G\'{o}mez-Camacho,
A. M. Moro, I. J. Thompson, and J. A. Tostevin,
Phys.\ Rev.\ C {\bf 80} (2009), 051601(R).


\bibitem{Matsumoto:2010mi}
T.~Matsumoto, K.~Kato, and M.~Yahiro,
Phys.\ Rev.\  C {\bf 82} (2010), 051602, 
arXiv:1006.0668 [nucl-th].


\bibitem{Nakamura}
    T.~Nakamura, \textit{et al}.,
    Phys. Rev. Lett. {\bf 103} (2009), 262501.

\bibitem{30Ne}
W. Horiuchi, Y. Suzuki, P. Capel, and D. Baye,
Phys. Rev. C \textbf{81} (2010), 024606.


\bibitem{Xu}
    H. M. Xu, {\it et al.},
    Phys. Rev. Lett. {\bf 73} (1994), 2027.

\bibitem{Eikonal-CDCC}
K.~Ogata, M.~Yahiro, Y.~Iseri, T.~Matsumoto, and M.~Kamimura,
Phys.\ Rev.\ C {\bf 68} (2003), 064609.


\bibitem{NNint}
B. Abu-Ibrahim, W. Horiuchi, A. Kohama, and Y. Suzuki,
Phys. Rev. C \textbf{77} (2008), 034607;
W. Horiuchi, Y. Suzuki, B. Abu-Ibrahim, and A. Kohama,
Phys. Rev. C \textbf{75} (2007), 044607.



\bibitem{Gogny}
J. Decharge and D. Gogny, Phys. Rev. C \textbf{21} (1980), 1568.

\bibitem{D1S}
J. F. Berger, M. Girod, and D. Gogny, Comp. Phys. Comm. \textbf{63} (1991), 1365.

\bibitem{Sn}
B. Jurado, \textit{et al}., Phys. Lett. B \textbf{649} (2007), 43.














\end{thebibliography}
\end{document}